\documentclass[]{spie}  

\usepackage{amsmath,amsfonts,amssymb}
\usepackage{graphicx}
\usepackage[colorlinks=true, allcolors=blue]{hyperref}
\usepackage{acronym,upgreek,gensymb}

\title{Performance estimates for spectrographs using photonic reformatters}

\author[a]{Robert. J. Harris}
\author[b]{Lucas Labadie}
\author[c]{Ulrike Lemke}
\author[d]{David G. MacLachlan}
\author[d]{Robert R. Thomson}
\author[a]{Sabine Reffert}
\author[a]{Andreas Quirrenbach}

\affil[a]{Zentrum f\"ur Astronomie der Universit\"at Heidelberg, Landessternwarte K\"onigstuhl, K\"onigstuhl 12, 69117 Heidelberg}
\affil[b]{I. Physikalisches Institut, Universit\"at zu K\"oln, Z\"ulpicher Str. 77, 50937 K\"oln, Germany}
\affil[c]{ESTEC, European Space Research and Technology Centre, Keplerlaan 1, 2201 AZ Noordwijk, Netherlands}
\affil[d]{SUPA, Institute of Photonics and Quantum Sciences, Heriot-Watt University, Edinburgh, EH14 4AS, UK}

\acrodef{AO}[AO]{Adaptive Optics}
\acrodef{FRD}[FRD]{Focal Ratio Degradation}
\acrodef{FWHM}[FWHM]{Full Width Half Maximum}
\acrodef{FRD}[FRD]{Focal Ratio Degradation}
\acrodef{NA}[NA]{Numerical Aperture}
\acrodef{PL}[PL]{photonic lantern}
\acrodef{PSF}[PSF]{Point Spread Function}
\acrodef{PIMMS}[PIMMS]{Photonic Integrated Multimode Spectrograph}
\acrodef{SN}[SN]{Signal to Noise}

\authorinfo{Further author information: (Send correspondence to R.J.H)\\R.J.H.: E-mail: rharris@lsw.uni-heidelberg.de, Telephone: +49 6221 541733}

\begin{document} 
\maketitle

\begin{abstract}
{Using a photonic reformatter to eliminate the effects of conventional modal noise could greatly improve the stability of a high resolution spectrograph. However the regimes where this advantage becomes clear are not yet defined. Here we will look at where modal noise becomes a problem in conventional high resolution spectroscopy and what impact photonic spectrographs could have. We will theoretically derive achievable radial velocity measurements to compare photonic instruments and conventional ones. We will discuss the theoretical and experimental investigations that will need to be undertaken to optimize and prove the photonic reformatting concept.}

\end{abstract}


\section{Introduction}

High resolution spectroscopy is one of the pillars of modern astronomy. When applied to stars it allows the derivation of temperature, stellar composition and velocity. It is also  is a fundamental technique for detecting exoplanets. 

However, as astronomers push for greater stability and resolution, the results are limited by new factors. Not only is the light from the star dispersed by larger amounts, meaning only brighter targets can be observed, but new noise limits are set by the instrument and star. Knowing the velocities of stars to the order of cm/s requires large telescopes, this means the spectrographs behind them have to grow in size, becoming more costly to manufacture. This also means controlling them to high precision becomes more difficult. To keep the possibility of mechanical flexure to a minimum, temperatures and pressures are highly controlled (e.g.\ HARPS is maintained at a pressure \textless $10^{-2}$ mbar and  a temperature of 17 \degree C, constant within 0.005 \degree C RMS\cite{2003Msngr.114...20M}) and this will only become more stringent in the next generation of instruments. Once the mechanical controls are in place, calibration requires devices like Iodine cells and Thorium Argon lamps, laser frequency combs \cite{steinmetz2008laser} and Fabry-Perot interferometers \cite{wildi2010fabry,schwab2015stabilizing}.



Once the above elements are controlled other problems arise, such as the stability of the input fibers to these spectrographs. It has been shown that conventional multimode fibers are subject to modal noise, \cite{goodman1981statistics,lemke2011modal} due to the wave nature of light producing modes within the fiber. If this is not properly removed, the resulting measured barycentre for a given wavelength will be altered (note this is different to  incomplete scrambling \cite{mccoy2012optical}), changing the resultant radial velocity measurement. 

Modern high resolution spectrographs use fiber scramblers, different core geometries and fiber agitators, to attempt to remove modal noise though not always completely. Modal noise also becomes worse at longer wavelengths, where scrambling becomes harder due to the inverse dependence of number of modes with wavelength.

The field of Astrophotonics \cite{Bland-Hawthorn2009} provides two potential solutions to the problem of modal noise. The first use a photonic lantern to seperate the modes and use agitation to effectively 'mix' them \cite{haynes2014new}. The second is to reformat the multimode fiber into lots of individual single modes, which can be dealt with separately. This can be done in many ways, \cite{harris2014comparison} such as \ac{PIMMS} \cite{Bland-Hawthorn2010}, TIGER \cite{2012arXiv1208.3006L} the photonic dicer \cite{2015MNRAS.450..428H}. Reformatting into a long slit, such as with the photonic dicer should in practise eliminate any barycentre shift (in one axis), however if the slit is imperfectly manufactured it could cause effects that have yet to be quantified and it is not yet shown whether using single modes may cause other problems (e.g. Ref.  \citenum{2015ApJ...814L..22H}).

Reformatting, also comes at a price (unless the input from the telescope is diffraction limited). All these extra modes must pass through the spectrograph and be sampled by the detector in order to retain throughput, increasing the number of pixels required for the spectrograph \cite{Harris2012,betters2013beating}. This can lead to uncompetitive sizes and costs in certain types of instruments \cite{2013MNRAS.428.3139H}. Because of this, science cases requiring a small telescope, \ac{AO}, space based instrumentation, a long wavelength and small field of view (few spaxels) \cite{Harris2012} are desirable. The next generation of high resolution spectrographs fit well into this category.

This paper will look at three existing high resolution spectrographs and identify if and where modal noise is a limiting factor. It will compare this to models of photonic spectrographs and identify where they can provide a competitive edge.

\section{Methodology}

The methodology for calculating the exposure time for a conventional spectrograph and photonic one are detailed here. Note that there are many assumptions in the model, these include:

\begin{enumerate}
\item All the models assume no optical aberrations due to telescope or the spectrograph.
\item The only sources of noise are due to the detector and photon statistics. No other sources of noise are included. This would mean a perfectly stable spectrograph.
\item We assume the coupling from the telescope and the spectrograph have no wavelength dependent response. This is of course not true in real spectrographs, but for our simple model will suffice.
\end{enumerate}

With the above assumptions, we can state that the \ac{SN} of both conventional and photonic models as

\begin{equation}
	SN = \frac{N_{\rm{star}}} {\sqrt{N_{\rm{star}} +  N_{\rm{sky}} + n_{y} N_{\rm{dark}} t + n_{y} N^{2}_{\rm{read}}}}.
	\label{eqn:s_n}
\end{equation}


Where $N_{\rm{star}}$ is the number of electrons produced on the detector from the source object, $N_{\rm{sky}}$ the number of electrons produced on the detector from the sky, $n_{\rm{y}}$ is the number of pixels per spectral resolution element, $N_{\rm{dark}}$ is the rate of electrons produced on the detector by dark current, $N_{\rm{read}}$ is the number of electrons produced from read noise and $t$ is the integration time. The rate of electrons from the star or sky for a given resolution element is

\begin{equation}
	N_{\rm{star/sky}} = \frac{F (\Delta \lambda) \epsilon A t }{P}
	\label{eqn:N_star}
\end{equation}

Where $F$ is the flux, $\Delta \lambda$ is the bin size, found by dividing the spectral resolution $\Delta \lambda = \lambda / R$ by the sampling, $\epsilon$ is the optical efficiency, $A$ is the area of the telescope and $P$ the energy of one photon.

Finally the flux is given by

\begin{equation}
	F =  z 10^{-0.4V}
	\label{eqn:flux}
\end{equation}

where $z$ is the zeropoint for the band (given by the ESO calculator) and $V$ is the apparent magnitude of the object.

The number of modes required to efficiently couple to the \ac{FWHM} of a \ac{PSF} can be approximated by

\begin{equation}
	M \approx \left( \frac{\pi \chi D_{T} }{4 \lambda} \right) ^{2}.
	\label{eqn:modes}
\end{equation}

Where $\chi$ is the seeing, $D_{\rm{T}}$ is the diameter of the telescope and $\lambda$ is the central wavelength. Note that this equation is for a circular fiber and does not account for the polarization within a single mode fiber. This is also for a fully illuminated fiber, whereas most spectrographs are fed at a slightly lower \ac{NA} (larger F\#) to account for \ac{FRD}, meaning not all the modes are initially illuminated.

\subsection{Signal to noise to radial velocity conversion}

To convert \ac{SN} to radial velocity we use the equation in Ref. \citenum{hatzes2010detection}. This is

\begin{equation}
	\sigma \mbox{(m/s)} = C (SN)^{-1} R^{-3/2} B^{-1/2} [f(Sp.T)]^{-1/2}  (v \sin i/2)^{-1}.
\end{equation}

Here $R$ is the spectral resolving power. $C$ is a constant of proportionality, which we take as $C$ $\approx$ 2.4 x $10^{11}$, to be the same as in Ref. \citenum{hatzes2010detection}. $B$ is the wavelength coverage used for the radial velocity measurement. The function $f(Sp.T)$ gives the relative line density as a function of stellar type, here we only take $f$=1 for a G-type star (in Ref. \citenum{hatzes2010detection} they also state for an A type star $f$= 0.1 and for an M type star $f$=10). Finally $v\sin i$ is the projected rotational velocity of the star in km/s, as with Ref. \citenum{hatzes2010detection} we take this as 2 km/s. As with other elements of this model we note that this does not fit all spectrographs and stars. For a more in depth discussion on calculating radial velocities we point the reader to Ref. \citenum{bouchy2001fundamental}.

\subsection{Conventional Spectrograph}

Here our conventional spectrograph will be limited either by photon noise, or by modal noise. Our equation for modal noise is taken from Ref. \citenum{lemke2011modal} and is given by

\begin{equation}
	{SN}_{\rm{coh}} = \rho \sqrt{\frac{M+1}{1-\rho^{2}}}.
\end{equation}

Where $\rho$ is the fractional area on the detector, as with Ref. \citenum{lemke2011modal} this equation is for coherent light. However, the light in a high resolution spectrograph is not totally coherent, due to the finite bandwidth of the spectral resolution element and the large number of modes populating the fibre. To allow for this we introduce a visibility term

\begin{equation}
	SN_{\rm{modal}} = \frac{1}{v}SN_{\rm{coh}}.
\end{equation}

Where $v$ is our visibility. This allows us to calculate our limiting \ac{SN} limit due to modal noise.

\subsection{Photonic Spectrograph} 

A photonic spectrograph using a spectral reformatter removes conventional modal noise within a high resolution spectrograph by putting all the modes from the \ac{PSF} into the spatial axis (the cross dispersed axis). This means when the spectrum is dispersed in the other axis that the output is stable. However this comes at a price, in order to adequately sample the reformatted slit extra detector pixels are needed for each spectral resolution element. 

We assume for simplicity that the modes are reformatted into one long line and that the modes are identical in size and shape. So each mode (spatial direction) can be sampled by two pixels, with a one pixel overlap. This means the number of pixels required in the spatial direction for a photonic spectrograph becomes

\begin{equation}
	n_y = M + 1.
\end{equation}

\section{Model Parameters}

In this section we fit our model to three high resolution fiber fed spectrographs. HARPS is a visible spectrograph at the 3.6m telescope in Chile. It is well known for having discovered many planets.

The second is the IR arm of CARMENES, one of the next generation of high resolution infrared spectrographs on 4m class telescopes. Recently commissioned on the 3.5m telescope on Calar Alto it ultimately hopes to achieve a precision of 1m/s. 

The third is Giano, an infrared spectrograph on the TNG telescope.  Also recently commissioned, this spectrograph has modal noise issues due to the longer wavelength it operates at.

The variable parameters used in the simulation are given in Tab.  \ref{tab:params}. We take the optical efficiency of the system to be 5.8\% and assume a magnitude 7 star in each case. We have exposure times of up to 20 minutes at 30 second intervals, as this represents close to saturation for the HARPS detectors. We take $\rho$ to be 0.9 (e.g. 90\% of the light from the fiber reaches the detector) and take three values for $v$, 0.973 represents an unagitated/scrambled fiber, 0.369 represents constant uniform agitation and 0.215 a random agitation (in this case by a PhD student)\cite{sturmer2014carmenes}. We also use the sky background as given by the ESO exposure time calculator.

\begin{table}[ht]
	\caption{Parameters for the modeled spectrographs.} 
	\label{tab:params}
	\begin{center}       
		\begin{tabular}{|c|c|c|c|} 
			\hline
			\rule[-1ex]{0pt}{3.5ex}   & HARPS & CARMENES (IR) & GIANO\\
			\hline\hline
			\rule[-1ex]{0pt}{3.5ex} $D_{\rm{T}}$ (m)& 3.6 & 3.5 & 3.58 \\
			\rule[-1ex]{0pt}{3.5ex} $\chi$ (arcseconds) & 1.  & 1.5  & 1 \\
			\rule[-1ex]{0pt}{3.5ex} $\lambda_{c}$ (nm) & 550  & 1335 & 1700 \\
			\rule[-1ex]{0pt}{3.5ex} Bandwidth (nm) &  310 & 750 & 1500  \\
			\rule[-1ex]{0pt}{3.5ex} Sky background & 21.7  & 14.4 & 14.4  \\
			\rule[-1ex]{0pt}{3.5ex} Sampling (pix) &  3.4 & 2.8 & 4  \\
			\rule[-1ex]{0pt}{3.5ex} Pixel size (${\upmu}$m)& 15  & 18 & 18  \\
			\rule[-1ex]{0pt}{3.5ex} Dark ($e^{-}$) & 0.0 & 0.01 & 0.05 \\
			\rule[-1ex]{0pt}{3.5ex} Read noise ($\rm{e}^{-} \rm{pix}^{-1}$) & 2.9  & 10 & 5 \\
			\rule[-1ex]{0pt}{3.5ex} $R$  & 100,000  & 80,400 & 50,000 \\		
			\hline
		\end{tabular}
	\end{center}
\end{table}





\section{Results}

\begin{figure}
\begin{center}
\begin{tabular}{c c}
\includegraphics[height=5.5cm]{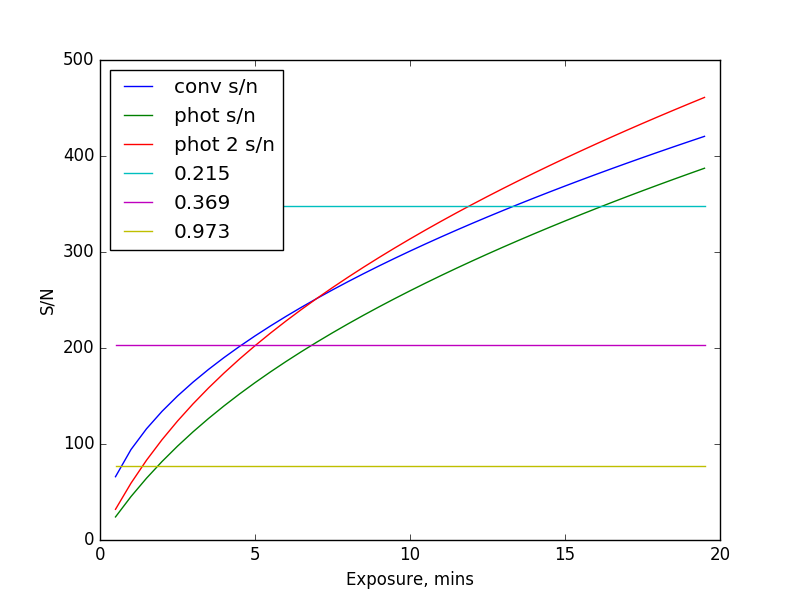}
\includegraphics[height=5.5cm]{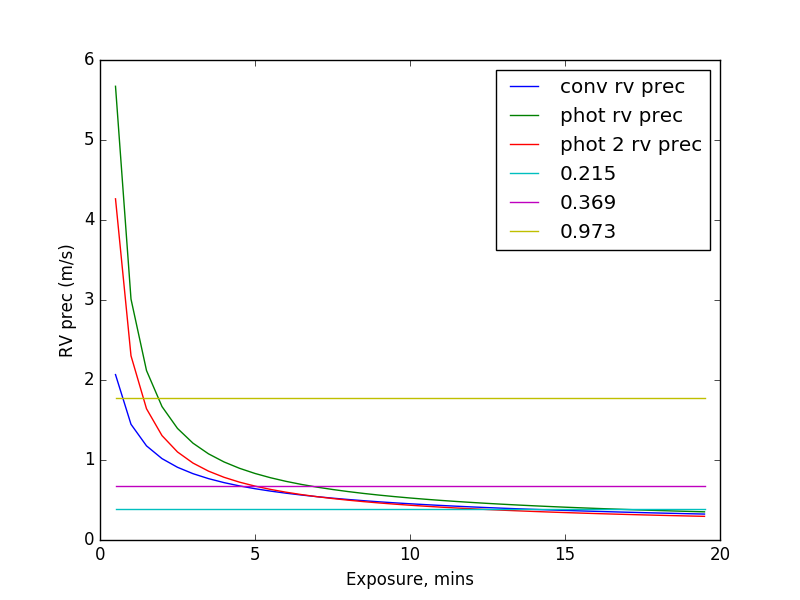}  
\\
(a) \hspace{6.7cm} (b)
\end{tabular}
\end{center}
\caption 
{ \label{fig:harps_results}
The graphs for (a) Signal to noise vs exposure time in minutes (b) Estimated radial velocity precision vs exposure time in minutes. The models are all based upon HARPS and represent a conventional (conv), photonic (phot) and photonic with lower sampling (phot 2) model. The three horizontal lines represent the modal noise limits (e.g. maximum achievable signal to noise or radial velocity precision) for a conventional spectrograph with no agitation/scrambling (0.973), mechanical agitation/scrambling (0.369) and random agitation/scrambling (0.215). Here you can see HARPS is not significantly limited by modal noise, so photonic technologies would not show improvement in achievable precision.} 
\end{figure}

\begin{figure}
\begin{center}
\begin{tabular}{c c}
\includegraphics[height=5.5cm]{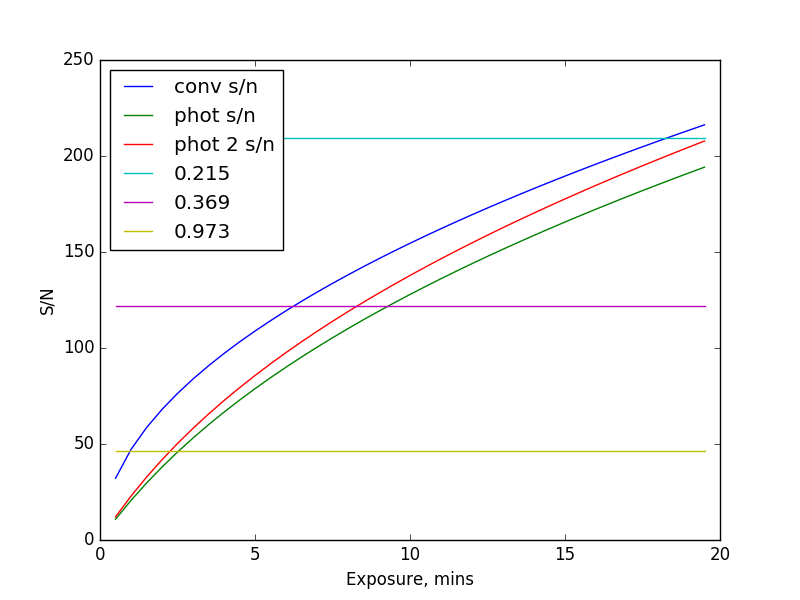}
\includegraphics[height=5.5cm]{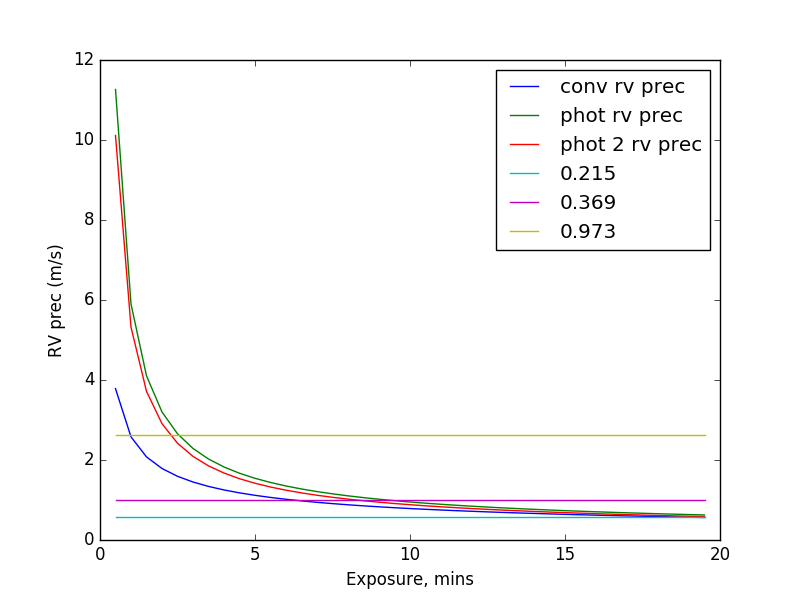}  
\\
(a) \hspace{6.7cm} (b)
\end{tabular}
\end{center}
\caption 
{ \label{fig:carmenes_results}
The graphs for (a) Signal to noise vs exposure time in minutes (b) Estimated radial velocity precision vs exposure time in minutes. The models are all based upon CARMENES and represent a conventional (conv), photonic (phot) and photonic  with lower sampling (phot 2) model. The three horizontal lines represent the modal noise limits (e.g. maximum achievable signal to noise or radial velocity precision) for a conventional spectrograph with no agitation/scrambling (0.973), mechanical agitation/scrambling (0.369) and random agitation/scrambling (0.215). Here you can see CARMENES is on the limit of being limited by modal noise, so photonic technologies may show some improvement in achievable precision.} 
\end{figure}

\begin{figure}
\begin{center}
\begin{tabular}{c c}
\includegraphics[height=5.5cm]{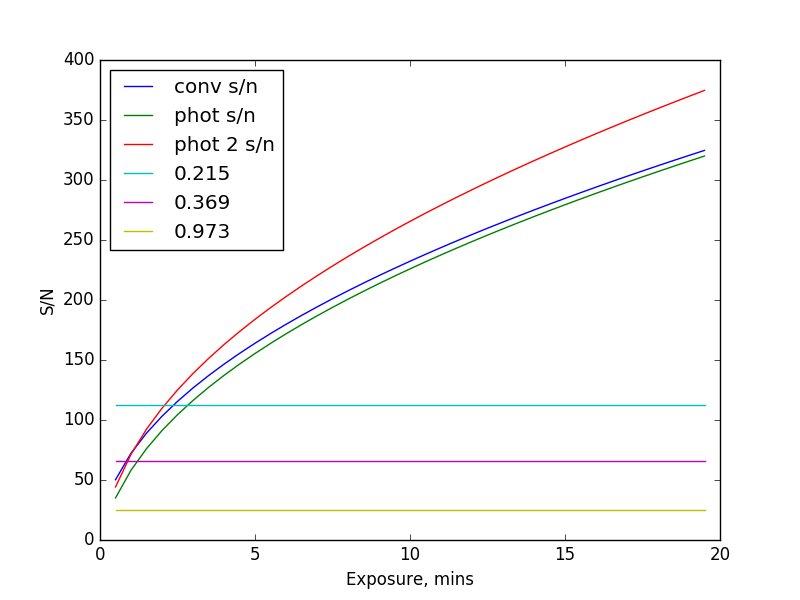}
\includegraphics[height=5.5cm]{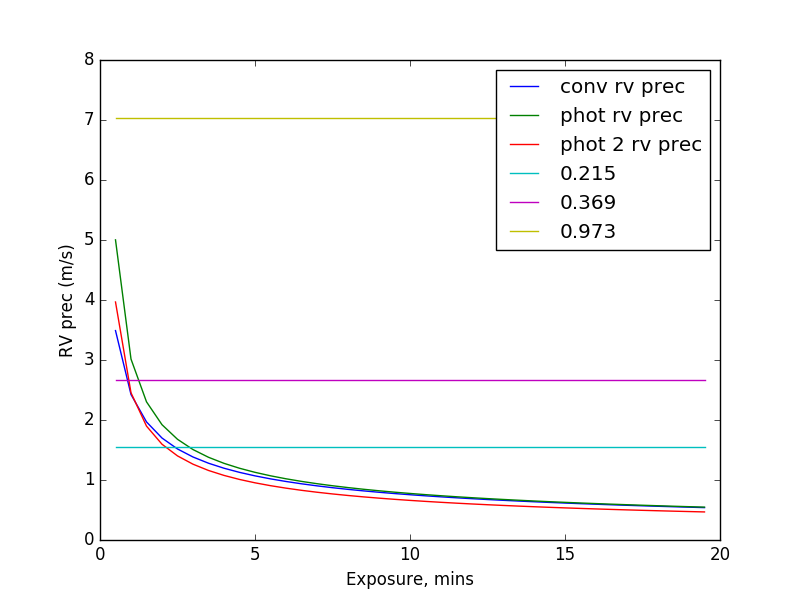}  
\\
(a) \hspace{6.7cm} (b)
\end{tabular}
\end{center}
\caption 
{ \label{fig:giano_results}
The graphs for (a) Signal to noise vs exposure time in minutes (b) Estimated radial velocity precision vs exposure time in minutes. The models are all based upon HARPS and represent a conventional (conv), photonic (phot) and photonic  with lower sampling (phot 2) model. The three horizontal lines represent the modal noise limits (e.g. maximum achievable signal to noise or radial velocity precision) for a conventional spectrograph with no agitation/scrambling (0.973), mechanical agitation/scrambling (0.369) and random agitation/scrambling (0.215). Here you can see that GIANO is limited by modal noise, so using photonic technologies would show improvement in achievable precision.} 
\end{figure}

We produce three models for each instrument. The first is based on a conventional spectrograph and the results for HARPS are matched to those in the ESO calculator and Ref. \citenum{hatzes2010detection}, in the graphs this is labeled as "conv". We then produce a photonic model, with the same sampling in the spectral direction, in the graphs this is "phot". Finally we produce a photonic model with a sampling set to 2.5 pixels per resolution element. We do this assuming the slit is a Gaussian with a constant width, so the sampling can be reduced, this is  "phot2". The results for each model are shown in Figs. \ref{fig:harps_results}, \ref{fig:carmenes_results} and \ref{fig:giano_results}.

Ignoring the limiting modal noise limits you can see from all the figures that the full photonic model "phot" achieves a lower \ac{SN} than a conventional spectrograph, due to the larger number of pixels required to sample the spectrum.
Reducing the sampling "phot2" means the \ac{SN} matches more closely for short exposures and for longer exposures results in a great signal to noise for HARPS and GIANO.

Once modal noise limits are included it can be seen from Fig.~\ref{fig:harps_results} that if the output from the fiber was not scrambled ($v$ = 0.973) the \ac{SN} would be limited to just under 100, which is obviously not the case for HARPS. With scrambling the limiting \ac{SN} is around 350, which allows a precision of around 40cm/s to be reached, this is below the current long term goal of HARPS of 0.6~m/s, showing that in its current setup, modal noise has little effect upon HARPS.

Fig. \ref{fig:carmenes_results} shows the results for CARMENES. The spectrograph operates at a longer wavelength and is more influenced by modal noise. However the \ac{SN} due to the light from the star is also lower.  This means at the 1m/s accuracy level that CARMENES hopes to achieve modal noise should not be a problem, provided the fiber output is sufficiently scrambled. However if CARMENES wished to achieve a higher radial velocity precision, this would need to be accounted for.

Finally Fig.~\ref{fig:giano_results} shows the results for GIANO. This spectrograph operates with the longest central wavelength, with the widest bandwidth and lowest spectral resolution. It can be seen from our conventional model that modal noise limits the \ac{SN} to around 25 without scrambling and around 60 with (which matches reasonably well to the quoted values of 20 and 50). This means the spectrograph is limited to around 1.5 m/s (as opposed to the current 7 m/s). If the modal noise was eliminated this could be pushed far lower, to sub m/s precision.

\section{Conclusions}

We have used toy models to investigate where modal noise could be a problem in high resolution spectroscopy. To do this we have compared conventional instrumentation to models of photonic spectrographs for HIRES, CARMENES and GIANO. As with other papers in the field we have concluded that modal noise is most problematic at longer wavelengths.

For all directly comparable models we found that photonic spectrographs would achieve lower signal to noise for a given exposure time due to the required number of modes (and hence pixels to sample them). This leads to a lower radial velocity precision when only using at photon statistics to calculate signal to noise (this is inline with previous results).
In cases where we reduced the sampling of the slit we found similar or in some cases higher signal to noise for photonic spectrographs. This should be taken with caution as we have assumed that the slit is uniform and no aberrations are introduced by the spectrograph.

When the limit due to modal noise is taken into account we found that HIRES and CARMENES should not be hugely affected (provided that the fiber is scrambled, which is the case for both), but that it was the limiting factor for GIANO (again consistent with results). If a photonic spectrograph can be shown to reduce or eliminate modal noise it would be very useful in this case.

Our toy model does have limitations and further work is needed. Summarized briefly:

\begin{enumerate}
\item While initial results are promising it still needs to be shown that photonic reformatters are modal noise free. This will need to be be explored both theoretically and experimentally. 

\item Even if the photonic reformatter can be shown to be modal noise free, other factors like polarization response (e.g Ref.  \citenum{2015ApJ...814L..22H}), may need to be taken into account. Investigations should be conducted to see how the response of the reformatter compares with conventional spectrographs.

\item Here we have assumed that each mode along the slit needs to be sampled by two pixels in the spatial direction. There have been suggestions before of reducing this by using cylindrical lenses or rectangular pixels. This should be investigated.

\end{enumerate}

Finally additional spectrographs can be modelled to see how they respond. To this end the code is freely available. It can be found at https://zenodo.org/record/54765.

\acknowledgments 

We would like to thank Jakob Vinther from the ESO ETC team for answering many questions on the ESO HARPS ETC model (https://www.eso.org/observing/etc/bin/gen/form?INS.NAME=HARPS+INS.MODE=spectro).  We would also like to thank Andr\'{e} Germeroth for his help and advice. 
Robert J. Harris is funded/supported by the Carl-Zeiss-Foundation

\bibliography{ref} 
\bibliographystyle{spiebib} 

\end{document}